\documentclass[aps,prl,twocolumn,superscriptaddress,groupedaddress]{revtex4}

\usepackage{amsmath}
\usepackage{amssymb}
\usepackage{graphicx}
\usepackage{epsfig}
\usepackage{dcolumn}
\usepackage{bm}
\usepackage{bm}
\usepackage{blindtext}
\usepackage{natbib}
\usepackage{booktabs}
\usepackage{mathrsfs}
\usepackage{epstopdf}
\usepackage{color}

\usepackage{textcomp}
\usepackage{times}
\usepackage{float}
\usepackage{latexsym,amsmath,amssymb,bm,euscript}
\usepackage{subfigure}
\usepackage[colorlinks=true,linkcolor=blue,citecolor=blue]{hyperref}
\usepackage{hyperref}
\usepackage{soul}
\usepackage[normalem]{ulem}
\usepackage{lettrine}
\usepackage{xspace}
\usepackage{textcomp}

\def\be{\begin{equation}}
\def\ee{\end{equation}}
\def\bea{\begin{eqnarray}}
\def\eea{\end{eqnarray}}

\begin{document}

\title{Spin dynamics of the $E_8$ particles}

\author{Xiao Wang}
\affiliation{Tsung-Dao Lee Institute,
Shanghai Jiao Tong University, Shanghai, 201210, China}

\author{Konrad Puzniak}
%\altaffiliation{konrad.puzniak@helmholtz-berlin.de}
\affiliation{\mbox{Helmholtz-Zentrum Berlin f\"{u}r Materialien und Energie GmbH, Hahn-Meitner Platz 1, D-14109 Berlin, Germany}}
\affiliation{\mbox{Institut f\"{u}r Festk\"{o}rperphysik, Technische Universit\"{a}t Berlin, Hardenbergstra{\ss}e 36, D-10623 Berlin, Germany}}

%\author{A A}
%\affiliation{\mbox{Helmholtz-Zentrum Berlin f\"{u}r Materialien und Energie GmbH, Hahn-Meitner Platz 1, D-14109 Berlin, Germany}}

\author{Karin Schmalzl}
\affiliation{\mbox{Jülich Centre for Neutron Science JCNS, Forschungszentrum Jülich GmbH, Outstation at ILL, 38042 Grenoble, France}}

\author{C. Balz}
\affiliation{ISIS Neutron and Muon Source, STFC Rutherford Appleton Laboratory, Didcot OX11 0QX, United Kingdom}

\author{M. Matsuda}
\affiliation{Neutron Scattering Division, Oak Ridge National Laboratory, Oak Ridge, Tennessee 37831, USA}

\author{Akira Okutani}
\affiliation{Center for Advanced High Magnetic Field Science, Graduate School of Science, Osaka University, Osaka 560-0043, Japan}

\author{M. Hagiwara}
\affiliation{Center for Advanced High Magnetic Field Science, Graduate School of Science, Osaka University, Osaka 560-0043, Japan}

\author{Jie Ma}
\altaffiliation{jma3@sjtu.edu.cn}
\affiliation{Tsung-Dao Lee Institute,
Shanghai Jiao Tong University, Shanghai, 201210, China}
\affiliation{School of Physics \& Astronomy, Shanghai Jiao Tong University, Shanghai, 200240, China}
\affiliation{Collaborative Innovation Center of Advanced Microstructures, Nanjing 210093, China}

\author{Jianda Wu}
\altaffiliation{wujd@sjtu.edu.cn}
\affiliation{Tsung-Dao Lee Institute,
Shanghai Jiao Tong University, Shanghai, 201210, China}
\affiliation{School of Physics \& Astronomy, Shanghai Jiao Tong University, Shanghai, 200240, China}
\affiliation{Shanghai Branch, Hefei National Laboratory, Shanghai 201315, China}

\author{Bella Lake}
\altaffiliation{bella.lake@helmholtz-berlin.de}
\affiliation{\mbox{Helmholtz-Zentrum Berlin f\"{u}r Materialien und Energie GmbH, Hahn-Meitner Platz 1, D-14109 Berlin, Germany}}
\affiliation{\mbox{Institut f\"{u}r Festk\"{o}rperphysik, Technische Universit\"{a}t Berlin, Hardenbergstra{\ss}e 36, D-10623 Berlin, Germany}}

\begin{abstract}
In this article, we report on inelastic neutron scattering measurements on a quasi-1D antiferromagnet BaCo$_2$V$_2$O$_8$ under a transverse magnetic field applied along the (0,1,0) direction. Combining results of inelastic neutron scattering experiments, analytical analysis, and numerical simulations, we precisely studied the $E_8$ excitations appearing in the whole Brillouin zone at $B_c^{1D}\approx 4.7$ T. The energy scan at $Q=(0,0,2)$ reveals a match between the data and the theoretical prediction of energies of multiple $E_8$ excitations. Furthermore, dispersions of the lightest three $E_8$ particles have been clearly observed, confirming the existence of the $E_8$ particles in BaCo$_2$V$_2$O$_8$. Our results lay down a concrete ground to systematically study the physics of the exotic $E_8$ particles.
\end{abstract}

\maketitle

Unlike the classical phase transition driven by thermal fluctuations, the quantum phase transition arises at zero temperature when the system is tuned by a non-thermal parameter~\cite{sachdev_2011}. For a continuous quantum phase transition, novel physics with higher symmetry may emerge at the quantum critical point (QCP), classified by a set of critical exponents manifested into scaling form~\cite{sachdev_2011}. Moreover, when the system is driven away from the QCP with a relevant perturbation, exotic physics may further emerge due to the strong renormalization of the almost infinite low-lying excitations,
which is ``emergence of emergence"~\cite{Zamolodchikov:1989fp,dorey1996,dorey1990}.

\begin{figure}[t]
\includegraphics[width=1\linewidth]{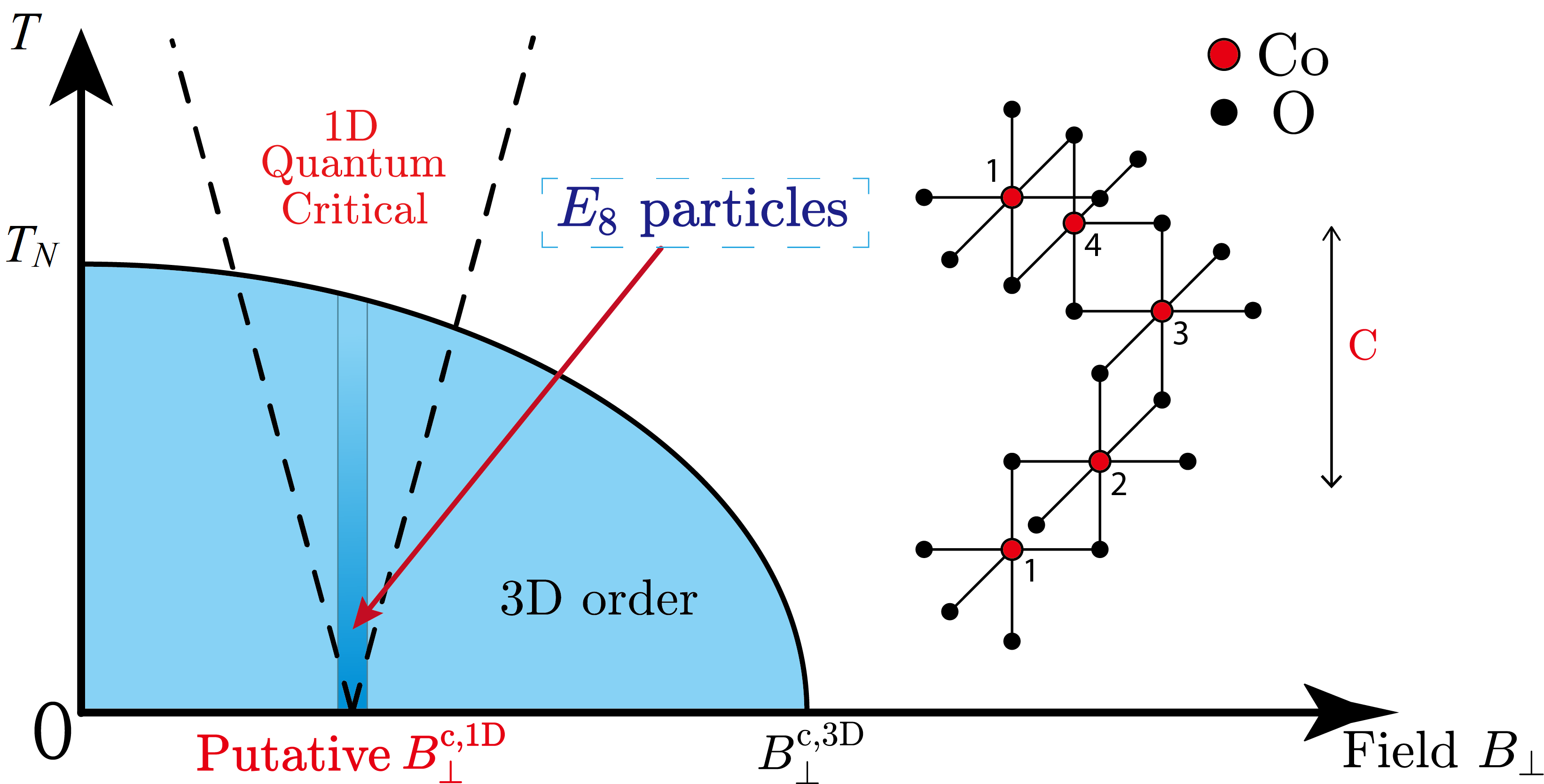}
\caption{Schematic phase diagram of BaCo$_2$V$_2$O$_8$ in a transverse magnetic field, light blue indicates the AFM order phase while the deep blue area covers the location of emergent exotic $E_8$ particles around the putative 1D QCP. The inset shows one of the CoO$_6$ screw chains.}
\label{fig:plot_phase}
\end{figure}

One such paradigmatic model is the transverse-field Ising
chain (TFIC)~\cite{sachdev_2011,Pfeuty}. When an Ising chain is tuned to its QCP by a magnetic field applied transverse to its Ising anisotropy, a central
charge 1/2 conformal field theory emerges with corresponding scaling exponents
falling into the class of Ising universality see Fig.~\ref{fig:plot_phase}~\cite{Pfeuty,Daniel}. Surprisingly,
when it is further perturbed by a longitudinal field parallel to the Ising direction,
the quantum $E_8$ integrable model emerges - a massive relativistic quantum field theory containing eight massive $E_8$ particles whose relative masses have precise values as listed in the first row of Table I. The physics of the model is described by scattering of the $E_8$ particles, which is characterized by the maximal exceptional Lie $E_8$ algebra~\cite{Zamolodchikov:1989fp,DELFINO1995724,Zou_2021,PhysRevB.101.220411,xiao_2021}.

For the experimental realization of the exotic $E_8$ physics,
two conditions need to be satisfied: accessing the Ising universality
and the presence of a small perturbation field along the Ising direction.
An early inelastic neutron scattering (INS) experiment on the ferromagnetic chain compound CoNb$_2$O$_6$ provided evidence of the existence of the lightest two particles:
the ratio of the energies of the lowest two peaks echos the
Golden ratio of the two lightest $E_8$ particles' masses~\cite{Coldea_2010}.
However, there is an apparent deviation in the spectrum continuum region between
the recent THz experiment on the material \cite{PhysRevB.101.220411} and
the analytical result \cite{xiao_2021}, which implies more efforts \cite{Armitage_2021, Fava25219} are needed for confirming the existence of the $E_8$ physics in the material CoNb$_2$O$_6$.

Recently, the quasi-1D Heisenberg-Ising antiferromagnetic (AFM) material BaCo$_2$V$_2$O$_8$ (a member of a family of materials with formula $AM_2$V$_2$O$_8$ in which: $A$ = Sr, Ba, and the magnetic ions are $M$ = Cu, Ni, Co, Mn), has attracted the attention of several experimental studies \cite{Faure:2017iup, PhysRevB.87.224413}.
When the materials are tuned by a transverse magnetic field, quantum criticality
with Ising universality is observed \cite{Zou_2021}.
Then the $E_8$ excitations with zero transfer momentum have been sought at low temperatures where the interchain coupling and long-range magnetic order provide the longitudinal field \cite{Zou_2021, PhysRevB.101.220411}. These $E_8$ excitations are measured by inelastic neutron scattering \cite{Zou_2021} (in this case all the $E_8$ particles were found) and terahertz spectroscopy \cite{PhysRevB.101.220411} (in the latter case the excitations
up to the fifth $E_8$ particles were found) and compared successfully to theory.

However, all these measurements were confined to \textit{only} the AFM zone center where the precise $E_8$ masses have already been calculated.
Furthermore, other peaks due to combinations of the $E_8$ particles and zone folding were also observed, which, while fully explainable, resulted in many overlapping excitations decreasing the certainty of the results. Considering that the $E_8$ model is a massive relativistic quantum field theory, a full investigation of the {\it relativistic dispersion} of the $E_8$ particles in the whole Brillouin zone is necessary for a complete realization of the $E_8$ physics in the material BaCo$_2$V$_2$O$_8$.

In this work, we combine and compare experimental, analytical, and numerical approaches for the dispersion of the lightest three branches of excitations and unambiguously demonstrate the existence of the exotic $E_8$ particles in the AFM material BaCo$_2$V$_2$O$_8$. BaCo$_2$V$_2$O$_8$
%is a well known 1D AFM spin-chain material, which 
crystallizes in tetragonal symmetry (space group $I4_1$/acd No. 142) with the lattice parameters: $a$ = $b$ = 12.40 \r{A} and $c$ = 8.375 \r{A} \cite{PhysRevB.87.224413}.
%BaCo$_2$V$_2$O$_8$ crystal has a complex magnetic structure.
The magnetic Co$^{2+}$ ions have effective spin of $S=\frac{1}{2}$ and are arranged in edge-sharing CoO$_6$ octahedra forming 4-fold screw chains running along the ${\bf c}$-axis
%, and separated by non-magnetic V$^{5+}$ and Ba$^{2+}$ ions
(see inset of Fig. 1).
There are four screw chains per unit cell, two of which rotate clockwise while the other two rotate anticlockwise \cite{PhysRevB.87.054408}. The Co$^{2+}$ ions are coupled by strong AFM interactions within the screw chains which have partial Ising (XXZ) anisotropy favouring spin directions parallel to the ${\bf c}$-axis.
%where the interaction strength is $J= 5.60$~meV and the anisotropy parameter is $\epsilon = 0.46$ forcing the spins to point in an up-down-up-down alignment along the ${\bf c}$-axis.
In the absence of an external magnetic field, BaCo$_2$V$_2$O$_8$ develops a long-range magnetic N\'eel order below $T_{\mathrm{N}}=5.5$~K due to weak interchain coupling, where neighboring spins are aligned antiferromagnetically along the screw chains and ferromagnetically (antiferromagnetically) between chains along the ${\bf a}$ (${\bf b}$) direction respectively. The spins point almost parallel to the ${\bf c}$-axis.

Under an external transverse magnetic field applied perpendicular to the spin direction (e.g.\ parallel to the ${\bf b}$-axis), BaCo$_2$V$_2$O$_8$ undergoes a three-dimensional (3D) quantum phase transition at $\mu_0 H_{\bot}^{c,3D} \approx 10.3$~T (Fig.~\ref{fig:plot_phase}) that was identified, by a combination of field theory, numerical analysis, and neutron scattering experiments, as a spin-flop transition from the ${\bf c}$ to ${\bf a}$ directions~\cite{Faure:2017iup}. This transition originates from the complex $g$-factor of BaCo$_2$V$_2$O$_8$. A one-dimensional (1D) quantum phase transition was also discovered at the lower transverse magnetic field of $\mu_0 H_{\bot}^{c,1D}= 4.7$~T~\cite{Weiqiang2019, Zou_2021}. This transition lies hidden within the dome of the 3D magnetic order (see Fig.~\ref{fig:plot_phase}) which provides the (staggered) longitudinal magnetic field required for the emergence of the $E_8$ quantum particles.

The aim of this paper is to investigate BaCo$_2$V$_2$O$_8$ at its putative critical transverse field $\mu_0 H_{\bot}^{c,1D} \approx 4.7$~T~\cite{Zou_2021} by combining an inelastic neutron scattering study of the dispersion of $E_8$ particles with a theoretical analysis based on the infinite time-evolving block decimation (iTEBD) technique~\cite{PhysRevLett.98.070201}. The plan of this paper is as follows: firstly we describe our experimental and theoretical approaches, then the neutron measurements of the dispersion of $E_8$ particles are presented and compared to the numerical simulations based on the iTEBD technique. Excellent agreement is achieved as a function of wavevector and energy for the lowest three $E_8$ particles.

%%%%%%%%%%%%%%%%%%%%%%%%%%%%%%%%%%%%%%

\begin{table*}[btp]
\caption{Predicted ratios of the energies of the single $E_8$ particle (first row) and the multi-particle (second row) excitations along with their expected values in BaCo$_2$V$_2$O$_8$ (third row). The experimental  excitation energies at various wavevectors from LET and IN12 are then listed along with the values obtained from iTEBD.}
\centering
\begin{tabular}{p{0.11\linewidth}p{0.065\linewidth}p{0.065\linewidth}p{0.065\linewidth}p{0.065\linewidth}p{0.065\linewidth}p{0.065\linewidth}p{0.065\linewidth}p{0.065\linewidth}p{0.065\linewidth}p{0.065\linewidth}p{0.065\linewidth}p{0.05\linewidth}}
\hline \hline
Single & $m_1$ & $m_2$ & $m_3$ & & $m_4$ & & $m_5$ & & & $m_6$ & & $m_7$ \\
\hline
Multi & & & & 2$m_1$ & & $m_1$+$m_2$ & &$m_1$+$m_3$ &3$m_1$ & &$2m_2$  \\
\hline \hline
Theoretical ratio $m_i$/$m_1$ & 1 & 1.618 & 1.989 & 2 & 2.405 & 2.618 & 2.956 & 2.989 & 3 & 3.218 & 3.236 & 3.891 \\
\hline
BaCo$_2$V$_2$O$_8$ theory [meV] & 1.26 & 2.04& 2.50  & 2.52 & 3.03 & 3.30 & 3.72 & 3.73 & 3.75 & 4.05 & 4.08  & 4.90\\
\hline
LET (0,0,2) [meV] & 1.28 & 2.04 & 2.52 & 2.58 & 3.04 & 3.30 & - & - & - & - & - & -\\
\hline
%LET (2,0,1) [meV] & 1.27 & 1.97 & 2.54 & - & - & - & - & - & - & - & - & -\\
%\hline
IN12 (0,0,2) [meV] & 1.26 & 2.05 & 2.49 & - & - & - & - & - & - & - & - & - \\
\hline
iTEBD (0,0,2) [meV] & 1.26  & 2.06  & 2.50  & 2.52  & 3.06  &  3.32  & 3.64  & 3.70  & 3.78  & 4.04  & 4.10  & 4.84\\
\hline
\end{tabular}
\end{table*}

\par

Two large, high-quality single crystals of BaCo$_2$V$_2$O$_8$ were grown using the floating-zone technique at Osaka University, Japan, and at the Core Lab for Quantum Materials, Helmholtz Zentrum Berlin f\"ur Materialien und Energie (HZB), Germany. Inelastic neutron scattering was performed to measure the magnetic excitations on the cold neutron multichopper spectrometer, LET (at the ISIS Facility, Rutherford Appleton Laboratory, UK) using the HZB crystal (mass 4.13 g) \cite{LET}. INS experiments were also performed on the cold neutron triple-axis spectrometer, IN12 from the Forschungszentrum Jülich Collaborating Research Group (FZJ-CRG) installed at Institut Laue Langevin (ILL) France, using the Osaka crystal (mass 3.66 g).

\begin{figure}[h]
\includegraphics[width=1.0\linewidth]{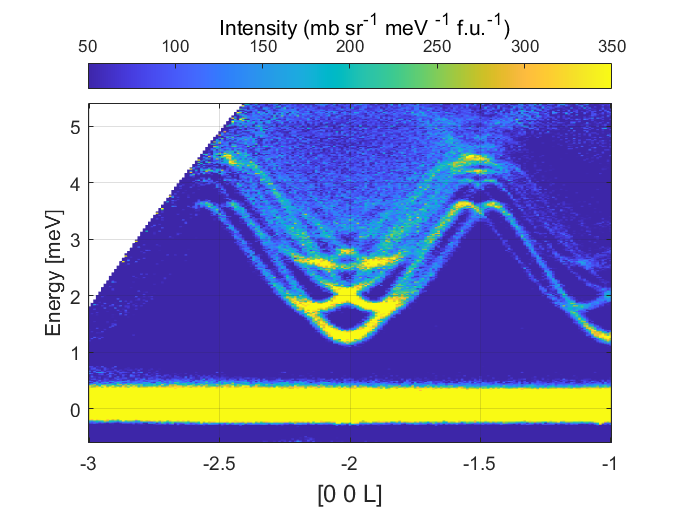}
\caption{Magnetic intensity of BaCo$_2$V$_2$O$_8$ measured in a transverse magnetic field of $\mu_0 H_{\bot}^{c,1D}= 4.7$~T at $T=0.3$~K using the LET spectrometer. The data is displayed in absolute units as a function of wavevector ${\bf Q}$ = (0,0,L) and energy, for neutron incident energy $E_i$ = 6 meV (integration range: $-1.0 \leq H \leq 1.0$ \& $-2 \leq K \leq 2$).
}
\label{fig:plot_maps}
\end{figure}

For the LET experiment, the single crystal was aligned in the (0,K,L) horizontal scattering plane and a vertical field cryomagnet was used to apply a constant magnetic field of $B=4.7$~T along the ${\bf a}$-axis to reach the 1D QCP. These measurements were carried out at $T=0.3$~K using a $^3$He-insert. This temperature is well below the N\'eel temperature ($T_{\mathrm{N}}=5.5$~K) ensuring the presence of the effective longitudinal perturbing field necessary to stabilize $E_8$ physics. Using repetition rate multiplication and the chopper frequencies 280/140 Hz, incident neutron energies of $E_i$ = 22.69, 13.21, 8.51, 6.00, 4.42, 3.42, 2.70 meV were achieved with corresponding elastic energy resolutions of $\Delta E$ = 0.91, 0.41, 0.22, 0.14, 0.094, 0.065, 0.048 meV. The INS data were processed using the MANTID and HORACE software packages and converted to absolute units. The spectrum for the incident energy of 6 meV is displayed in Fig.~\ref{fig:plot_maps} as a function of energy and wavevector transfer along the chain direction, (0,0,L).

\begin{figure*}
\includegraphics[width=1.0\linewidth]{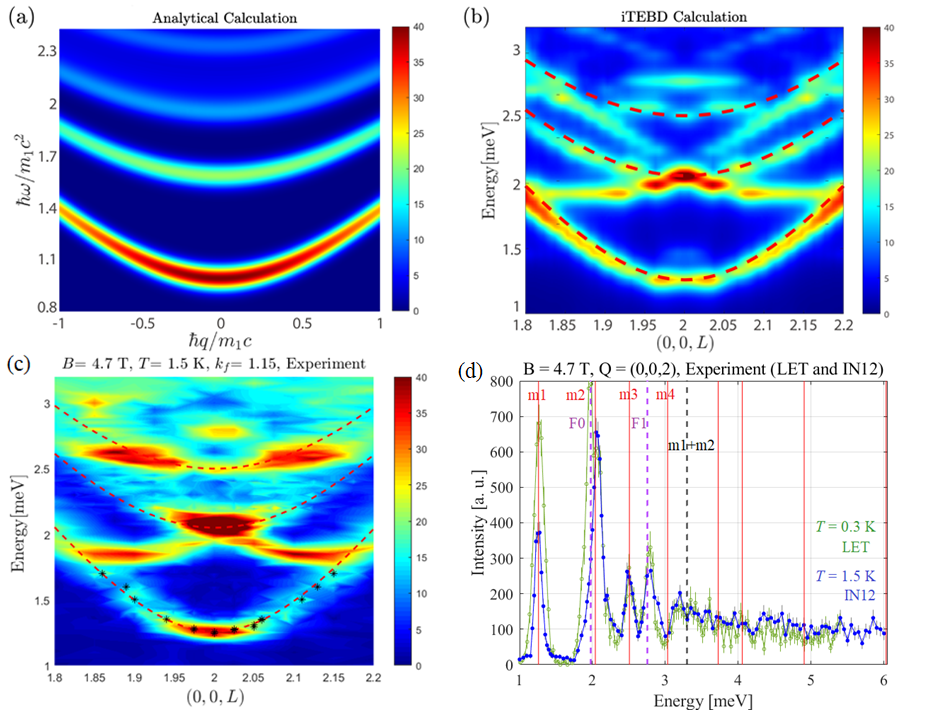}
\caption{(a) Analytical calculation of dispersion of three lightest $E_8$ particles. (b) Numerical simulation of dispersion using the iTEBD method, based on Eq.~(\ref{eq:DSF}). The dashed lines illustrate dispersions of the three lightest $E_8$ particles given by Eq.~(\ref{eq:fit}). (c) Energy-wavevector map of the magnetic excitations constructed from constant-energy and constant-wavevector scans measured on the IN12 spectrometer at $B^{1D}_c$ = 4.7 T and $T = 1.5$~K. The black symbols are the peak positions extracted from fitting the individual scans, whereas the dashed lines are fits to the dispersions to the lowest three $E_8$ particles using Eq.~(\ref{eq:fit}). (d) shows an energy scan measured at the wavevector $Q$ = (0,0,2) on LET and IN12, with the theoretically predicted energies of the first $E_8$ excitations indicated by the red solid lines. Identified peaks are labelled $m_n$ (single $E_8$ excitations), $m_n+m_m$ (multi-$E_8$ excitations) and $F_n$ (zone-folding peaks).}
\label{fig:plot1}
\end{figure*}

For the IN12 experiment, the crystal was aligned with the ${\bf a}$- and ${\bf c}$-axes within the horizontal instrumental scattering plane and a vertical DC magnetic field of 4.7 T was applied parallel to the ${\bf b}$-axis. A fixed final wavevector of $k_f = 1.15$~\AA$^{-1}$ was used, giving an energy resolution of $\Delta E \approx 0.114$~meV and wavevector resolution of $\approx 0.067$ r.l.u. A Beryllium filter was used to suppress higher-order wavelengths and spurious scattering. A series of energy scans at constant wavevector in the range from ${\bf Q}$ = (0,0,1.5) to (0,0,2.5) were performed over the energy range from $E = 0 5$~meV with steps of at least 0.05~meV. Constant-energy scans were also performed within this range. These measurements took place at a temperature of $T = 1.50$~K($\ll T_{\mathrm{N}}$). The constant-energy and constant-wavevector scans are combined together to make the energy-wavevector map in Fig.~\ref{fig:plot1} (c).

%\subsection{Theoretical model and dispersion of $E_8$ particles}

When applying a transverse field along (0,1,0) direction, the effective Hamiltonian for BaCo$_2$V$_2$O$_8$ is described by a 1D spin-1/2 Heisenberg-Ising model~\cite{Kimura_2013,Weiqiang2019,Zou_2021,PhysRevB.101.220411,Zou_2019}:
\begin{equation}
\begin{aligned}
 \mathcal{H} &=H_{XXZ}+H_{t}+H_{s}\\
H_{XXZ}&= J\sum_{n}[S^z_{n}S^z_{n+1}+\epsilon(S^x_{n}S^x_{n+1}+S^y_{n}S^y_{n+1})]\\
H_{t}&=-\mu_{B}g_{yy}H\sum_{n}[S^y_{n}+h_{x}(-1)^{n}S^{x}_{n}\\
& \;\;\;\; +h_{z}\cos(\pi\frac{2n-1}{4})S^z_{n}]\\
H_{s}&=-\mu_{B}H'\sum_{n}(-1)^n S^{z}_{n}
\label{eq:Hamil}
\end{aligned}
\end{equation}
where $S^{\alpha}_{n} = \frac{1}{2}\sigma^{\alpha}_{n}$($\alpha=x,y,z$) are spin-1/2 operators at site $n$ with Pauli matrices $\sigma^{\alpha}$. $J=5.8\;$meV, $\epsilon=0.46$, $h_{x(z)}=0.4(0.14)$, $g_{yy}=2.75$. The applied transverse field is set $\mu_{0}H=4.7\;$T, which is the critical field of the putative 1D QCP~\cite{Zou_2019,Zou_2021}. The effective staggered longitudinal field $\mu_{B}H'=0.018J$ comes from a mean-field treatment of the inter-chain coupling in the 3D ordering region below $T_{\mathrm{N}}$~\cite{Faure:2017iup,Zou_2021}.
 $H_s$ provides a necessary relevant perturbation for realizing the quantum $E_8$ physics \cite{Zou_2021}. Focusing on the parameter region around the putative 1D QCP, in the scaling limit, the effective Hamiltonian of the spin chain becomes~\cite{DELFINO1995724,Zou_2021,xiao_2021}
\begin{align}
\mathcal{H}_{E_8}=\mathcal{H}_{c=1/2}+h\int dx \sigma(x).
\label{eq:E8Hamil}
\end{align}
$\mathcal{H}_{c=1/2}$ is the Hamiltonian for a central charge 1/2 conformal
field theory, which describes the quantum critical physics of the TFIC.
$h$ and $\sigma (x)$ corresponding to the scaling limits of $\mu_{B}H'$ and $\sigma_j^z$
are the strengths of the perturbation field and the relevant primary field, respectively.

To determine the dispersion of $E_8$ particles and compare with the spectrum measured by INS, we calculate the spin dynamic structure factor (DSF) in the field theory frame, $D^{\alpha\alpha}(\omega,q)=\sum_{n=1}^{\infty}\frac{(2\pi)^2}{\prod_{a_{i}=1}^{8} n_{a_{i}}!}\int_{-\infty}^{\infty}\prod_{j=1}^{n}\frac{d\theta_{j}}{2\pi}|\langle 0|\sigma^{\alpha}|A_{a_{1}}(\theta_{1})...A_{a_{n}}(\theta_{n})\rangle|^{2}$\\
$\delta(\omega-\sum_{j=1}^{n}E_{j})\delta(q-\sum_{j=1}^{n}P_{j}),$
where $\alpha=x,z$, and $a_{i}=1...8$ are quasi-particles obtained from the quantum $E_{8}$ integrable theory~\cite{Zamolodchikov:1989fp,DELFINO1995724,Zou_2021,xiao_2021}. $n_{a_{i}}$ is the number of particle $a_{i}$ involved in the corresponding channel. $E_{j}=m_{a_{j}}\cosh\theta_{j}$ and $P_{j}=m_{a_{j}}\sinh\theta_{j}$ are the energy and momentum of particle $a_{j}$ in terms of the rapidity $\theta$. The two Dirac $\delta$-functions reflect the energy and momentum conservation of the scattering. The DSF of $\sigma^{x,z}$ can be directly calculated from quantum $E_{8}$ integrable field theory~\cite{xiao_2021}, and the DSF of $\sigma^{y}$ can be obtained from DSF of $\sigma^{z}$~\cite{PhysRevLett.113.247201}. The analytical result for the dispersion of the lightest three $E_8$ particles is shown in Fig.~\ref{fig:plot1}(a).
For a better comparison of the theoretical prediction from quantum $E_8$ field theory with the INS experimental result, two subtle issues are worth noting.

1. In the above field theory frame calculation, the speed of light is set as $c=1$. For the quantum $E_8$ model, as a massive relativistic quantum field theory, the dispersion of the $E_8$ particles follows the massive relativistic dispersion $E_i^2 = \Delta _i^2 + p_i^2{c^2}$, where $\Delta_i = m_i c^2$ and $p_i = m_i c$ with the ``rest mass" of the $i^{th}$ $E_8$ particle $m_i$ and the ``speed of light" $c$. When coming to a real material which actually is a lattice discrete in space, we need to re-scale the dispersion of the $E_8$ particles with the proper energy scale and length (momentum) scale serving as IR cutoffs. The theoretically expected energy peak corresponding to the lightest $E_{8}$ particle $m_{1}$ can be estimated from $E_{m_{1}}^{\text{theory}}=C_{\text{lattice}}H'^{8/15} \approx 1.2 \, \text{meV}$
\cite{Yang_2023}, where $C_{\text{lattice}} = 4.010\cdots$ is a modified constant for the lattice which originally comes from quantum $E_{8}$ field theory~\cite{E8_lattice}.
The value of $E_{m_{1}}^{\text{iTEBD}}$ matches the minimum gap
$\Delta_1 = 1.26 \,\text{meV}$ observed at the zone center (corresponding to zero transfer momentum), thus $\Delta_1$ can naturally serve as IR cutoff of the energy scale for the experimental data.
Since $\Delta_1 = m_1 c^2$ then we can pick up the corresponding IR momentum cutoff $p_1 = m_1 c$. By applying these two IR cutoffs scales we arrive at
\begin{equation}
\frac{{E_i^2}}{{{\Delta_{1}^{2}}}} = \frac{{\Delta _i^2}}{{\Delta_{1}^{2}}} + \frac{{p_i^2}}{(\Delta_{1}/c)^{2}}= \frac{{\Delta_i^2}}{{{{\Delta_1}^2}}} + \left( {\frac{{\hbar (L - 2)\pi /2d}}{{{m_1}c}}} \right)^2,
\label{eq:fit}
\end{equation}
where $p_i=\hbar(L - 2)\pi /2d$ is the momentum transfer with respect to ${\bf Q}$ = (0,0,2) and $d=8.4192/4=2.105$ is the nearest neighbor distance between Co$^{2+}$ ions projected onto the chain direction.
We need to determine the value of $c$ to obtain the IR cutoff for the momentum, whose value cannot be uniquely determined by the analytical theory but actually depends on the microscopic details of the material.

2. The four-fold periodicity of BaCo$_2$V$_2$O$_8$ leads to the sizable zone-folding effect of the experimental measurement, which makes the $E_{8}$ particles' dispersion shadowed by additional spectra. Such an effect cannot be obtained from the field theory calculation, instead, we need to go back to the original effective lattice model. By comparing spectra obtained from the lattice model and the field theory, the $E_{8}$ particles' dispersion will be extracted.

To make these two subtle issues clear, we carry out iTEBD simulation 
for the effective Hamiltonian Eq.~(\ref{eq:Hamil})
with $J=1$, $\epsilon=0.47$, and critical field $\mu_B g_{yy}H=0.15$~\cite{Zou_2019,Zou_2021}, 
\begin{align}
\begin{split}
D_{\rm{lat}}^{\alpha\alpha}(\omega,q)=\frac{1}{N}&\sum_{j,j'=1}^{N}\exp\{-iq(j-j')\}\\
&\times\int_{-\infty}^{\infty}dt \exp(i\omega t)\langle S_{j}^{\alpha}(t)S_{j'}^{\alpha}(0)\rangle,
\label{eq:DSF}
\end{split}
\end{align}
with total number of lattice sites $N$ ($N \to \infty$ in iTEBD), and spin-1/2 operators $S^{\alpha},\alpha=x,y,z$. 
The iTEBD simulation result is shown in Fig.~\ref{fig:plot1} (b), where
the value of the "speed of light" is found to be $c \approx (1.441 \pm 0.096)\times 10^{3}~m/s$ and the zone-folding effect can be identified as well.

The procedure of the iTEBD calculation is as follows: 1. Generate a four-periodic
ground state wave function of the effective Hamiltonian Eq.~(\ref{eq:Hamil})
with the parameters. The imaginary time-evolution is done
with fifth-order Trotter-Suzuki decomposition~\cite{Hatano2005}, where the imaginary time
slide is set as $d\tau=0.01$. The convergence condition is chosen as the
difference of the norm of singular values in the matrix product states being smaller than $10^{-12}$. The truncated dimension is chosen as $\chi=45$~\cite{PhysRevLett.98.070201,PhysRevA.79.043601}. 2. Calculate the DSF [Eq.~(\ref{eq:DSF})]
for $S^{x}$ and $S^{z}$, while the DSF of $S^{y}$ can be obtained from DSF of $S^{z}$ by using $D^{yy}(\omega,q)=\omega^2 D^{zz}(\omega,q)/(4J^2)$~\cite{PhysRevLett.113.247201}. For calculating this DSF with iTEBD algorithm, we first do real time and space propagation in Heisenberg picture, then by using Fourier transformation
we transform the obtained result into momentum and energy space
to obtain the final spectrum. The real time evolution is done
by a second order Trotter-Suzuki decomposition with $t=200,~dt=0.02$ for obtaining a relatively high accurate result near the Brillouin zone center. 3. A zone-folding effect is necessary to consider when obtaining the final spectrum for comparison with INS experimental results.

\par

The magnetic excitations of BaCo$_2$V$_2$O$_8$, measured using the LET time-of-flight spectrometer at $T$ = 0.3 K and $\mu_0 H = \mu_0 H_{\bot}^{c,1D}( = 4.7$~T) with incident neutron energy of 6 meV, is shown in Fig. 2 as a function of energy and wavevector transfer along the chain direction. To complement this, Fig.\ 3 (c) shows the low-energy excitations built from the energy- and wavevector-scans measured at $T=1.5$~K and $\mu_0 H = 4.7$~T on the IN12 spectrometer. An incredibly rich series of modes is found with complex dispersions and intensity modulations. First of all, the 4-fold screw-chain structure about the \textbf{c}-axis gives rise to four independent chain spectra shifted consecutively in wavevector along the chain by $\Delta L = 1$ r.l.u. Each individual spectrum is periodic over an interval of $\Delta L = 4$ r.l.u. due to the fact that there are four Co$^{2+}$ ions along each chain per unit cell (because of the screw-chain structure). Together these four spectra ensure that an antiferromagnetic zone center where the excitations are minimum is found at every integer value of $L$. At low energies, each spectrum is expected to consist of the $E_8$-particles observed as sharp (resolution-limited) gapped modes. In addition, multi-particle excitations are expected, due to the simultaneous creation of two or more $E_8$ particles such as $m_1+m_2$. These excitations form a continuum with a sharp lower boundary. Finally, zone-folding modes are expected which are a consequence of the screw chain structure of BaCo$_2$V$_2$O$_8$. Unlike the other excitations, these can be incommensurate with minima that do not always occur at integer values of $L$.

\begin{figure*}
\includegraphics[width=1.0\linewidth]{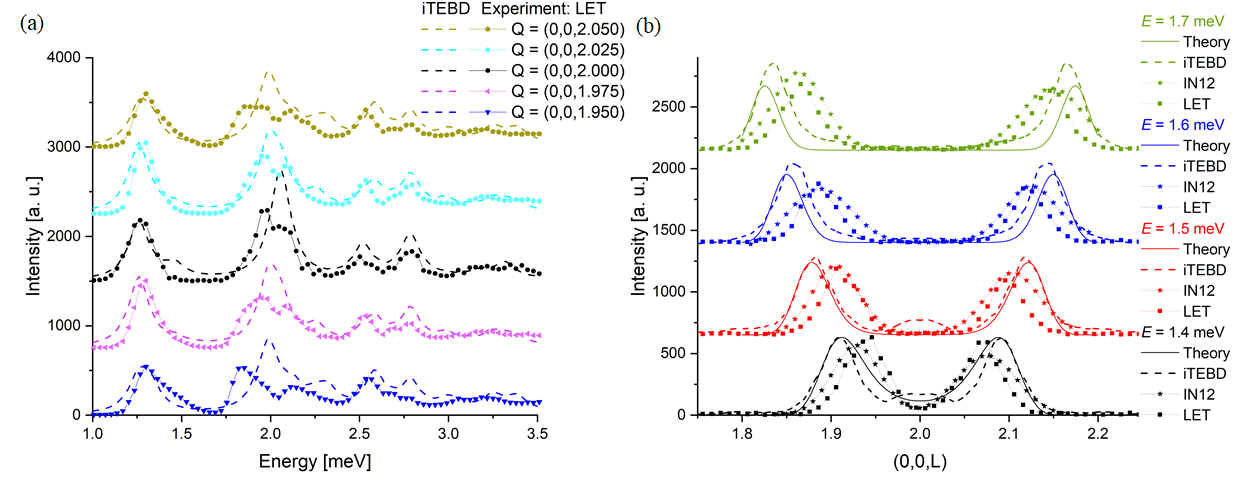}
\caption{(a) Comparison of DSF results between the iTEBD and the LET experimental data for constant momentum cuts. (b) Comparison of DSF results between the iTEBD data, the analytical data, and the experimental IN12 and LET data for constant energy cuts. The presented LET experimental constant momentum and constant energy cuts are shifted by an offset of 0.004 r.l.u (the offset was caused by experimental conditions). All DSF intensities are normalized up to the maximum intensity of experimental data.}
\label{fig:scans}
\end{figure*}

According to theory, the ratios of the energies of the eight $E_8$-particles have precise values \cite{borthwick2011did}. These values are given in the first row of Table I along with the expected values of the multiparticle excitations. The excitations were indeed observed previously at these energies at the antiferromagnetic zone centers (integer $L$-wavevectors) of BaCo$_2$V$_2$O$_8$ \cite{Zou_2021}. Here we reconfirm this observation using our higher resolution (but lower intensity) data. The energy scans through our LET and IN12 datasets are shown in Fig.~\ref{fig:plot1}(d). For both datasets, the first peak is observed at 1.26~meV. Multiplying the $E_8$ particle ratios by this value gives the theoretically expected peak positions for BaCo$_2$V$_2$O$_8$ (second row of Table I) which are also indicated by the solid vertical red lines in Fig.~\ref{fig:plot1}(d). The first three peaks in the LET and IN12 experimental data (at $m_1 = 1.26$, $m_2 = 2.04$, $m_3 = 2.52$~meV and at $m_1 = 1.26$, $m_2 = 2.05$, $m_3 = 2.49$~meV respectively) clearly lie at the expected positions of the first three $E_8$ peaks, (it should be noted that the multi-particle peak $2m_1$ coincides with $m_3$). It should be also noted that close to $m_2$ there is a zone-folding peak of 1.98~meV (indicated by F0). There exists also another zone-folding peak at 2.75~meV (indicated by F1). The fourth $E_8$ excitation at $m_4 = 3.04$~meV is very weak, while the feature at 3.30~meV is the lower boundary of the $m_1+m_2$ multi-particle continuum. At higher energies, it is difficult to distinguish the $E_8$ peaks due to overlapping continuua and zone-folding modes. The positions of peaks were found by fitting Gaussians. They are listed in Table I and are in agreement with the results of H. Zou {\it et. al.} \cite{Zou_2021}. Because we collected high-precision data over a wide range of wavevectors rather than at just the antiferromagnetic zone centers, we now have the opportunity to observe the behavior of the $E_8$-particles as well as the other excitations as a function of wavevector as well as energy. Returning to the energy-wavevector plots in Fig. 2 and 3(c), it is clear that the three lowest $E_8$ excitations actually form dispersive modes with parabolic curvature and a minimum at ${\bf Q}$ = (0,0,2). The zone-folding modes are now easily identified, such as the dispersive excitations which have minima at the incommensurate wavevectors (0,0,1.875) and (0,0,2.15) at $E$ = 1.9~meV and overlap with $m_2$ at (0,0,2). Finally above 3~meV broad diffuse scattering is observed due to the multi-particle continua and overlapping modes. The dispersions of the $E_8$ particles are expected to follow the theoretical expression given by Eq.~(\ref{eq:fit}), which can be modified as

\begin{equation}
%E_i=\sqrt{\Delta_i^2+\left(\frac{\hbar\pi c\cdot(L-2)}{2d}\right)^2}
E_i=\sqrt{\Delta_i^2+\left(\gamma\cdot(L-2)\right)^2},
\label{eu_eqn}
\end{equation}
where the single parameter $\gamma=\frac{\hbar \pi c}{2d}$ can be extracted. We simultaneously fit the lowest three $E_8$ dispersions and get the value $\gamma=8.07$~meV. The fitted dispersions are given by the red lines in Fig.~\ref{fig:plot1} (c) and show good agreement with the data. The `speed of light' was extracted from $\gamma$ and found to be $c \approx (1.643 \pm 0.041)\times 10^{3}~m/s$ (using $d=2.105$~\AA\ - the projection of the nearest neighbor Co$^{2+}$-Co$^{2+}$ distance onto the {\bf c}-axis). The value of $\gamma$, found from fitting the lowest three $E_8$ dispersions in the case of LET data, is $\gamma=8.81$~meV and the `speed of light' is found to be $c \approx (1.794 \pm 0.008)\times 10^{3}~m/s$. This value agrees well with the value of $c \approx (1.441 \pm 0.096)\times 10^{3}~m/s$ found from iTEBD. 

We further compare experimental, analytical, and numerical data with 
constant momentum scans [Fig.~\ref{fig:scans}(a)] and constant energy scans
[Fig.~\ref{fig:scans} (b)]. For the former comparison in Fig.~\ref{fig:scans}(a) 
the analytical data are not included due to the presence of zone-folding effects for energies above 2 meV. The iTEBD data agree very well with the experimental data. For the latter comparison Fig.~\ref{fig:scans}(b) in order to avoid mixture from zone-folding effect the energy window is chosen to match with the dispersion of the lightest $E_8$ particle.
The analytical and iTEBD data show excellent agreement 
with each other, and both show good agreement with the experimental data. We note
that there is about $1\%$ deviation in momentum for the peak position
corresponding to $1\%$ $m_1$ energy shift. This possibly is because the transverse field, applied during the experiment, is slightly smaller than the exact critical field which can result in
slightly heavier $E_8$ particles which implies a slightly larger
minimum gap $m_1$. $1\%$ $m_1$ shift corresponds to about 0.1 T to
0.2 T shift from the exact critical field whose value lies in the range of the identified
putative QCP $4.7 \pm 0.3$ T \cite{Zou_2021}.

%\section{Conclusions}
\par

To conclude, the quasi-1D antiferromagnet BaCo$_2$V$_2$O$_8$ is a very important material in the field of quantum magnetism. Among its unique properties are: Ising-like anisotropy, large intrachain versus weak but non-negligible interchain interactions, an anisotropic $g$ tensor producing easy-axis anisotropy, and effective staggered fields under the application of an external magnetic field. Combining the results of INS experiments, theoretical, and numerical iTEBD simulations, we have precisely studied the $E_8$ excitation spectrum appearing at the one-dimensional quantum critical point of $B^{1D}_c$ = 4.7 T. The observation of dynamical spectra through INS, together with excellent agreement with analytical analysis and iTEBD numerical simulations, enabled us to observe the dispersion of the first three $E_8$ particles and several multi-particles modes, paving the way toward possible manipulation of the $E_8$ particles.

%\section{Acknowledgments}
\par

The crystal growth and characterization took place at the Core Laboratory
Quantum Materials, Helmholtz Zentrum Berlin f\"ur Materialien und Energie,
Germany and at the Center for Advanced High Magnetic Field Science,
Graduate School of Science, Osaka University. The authors would like to thank ISIS and ILL facilities for the allocation of neutron beam-time. We would like to also thank C. Fritsche for his help in the preparation of the sample holder. K. P. is also very grateful to Dr. C. Rohr for all his important suggestions about the general structure of the article. The LET data (\url{https://doi.org/10.5286/ISIS.E.RB2210086}) were reduced using Mantid and were analyzed using the Horace-MATLAB software package. This work has, in part, been supported by National Natural Science Foundation of China No. U2032213 (J. M.), 12274288 (X. W. and J. W.) and the Innovation Program for Quantum Science and Technology Grant No. 2021ZD0301900 (X. W. and J. W.), 2022YFA1402702 (J. M.), and the Natural Science Foundation of Shanghai with grant No. 20ZR1428400 and Shanghai Pujiang Program with grant No. 20PJ1408100 (X.W. and J.W.), and Grants-in-Aid for  Scientific Research (Nos. 25220803 and 24244059) from MEXT. 

\par

X. W. and K. P contributed equally to this study. J. W., J. M., and B. L. conceived and coordinate the project. J. M., C. B, and B. L. designed the experiment.
X. W. and J. W. carry out analytical and iTEBD calculations and provide theoretical analysis. X. W., K. P., J. M., J. W., and B. L. wrote the manuscript. 

\bibliography{cite}
\bibliographystyle{apsrev4-2}

%%%% supp

\newpage
\appendix
\onecolumngrid
\section*{{\Large Supplemental Material---Spin dynamics of the $E_8$ particles}}

\section{Details of neutron experiments}

Two large, high-quality single crystals of BaCo$_2$V$_2$O$_8$ were grown using the floating-zone technique at Osaka University, Japan, and at the Core Lab for Quantum Materials, Helmholtz Zentrum Berlin f\"ur Materialien und Energie (HZB), Germany. Inelastic neutron scattering was performed to measure the magnetic excitations on the cold neutron multichopper spectrometer, LET (at the ISIS Facility, Rutherford Appleton Laboratory, UK) using the HZB crystal (mass 4.13 g) \cite{LET}. INS experiments were also performed on the cold neutron triple-axis spectrometer, IN12 from the Forschungszentrum Jülich Collaborating Research Group (FZJ-CRG) installed at Institut Laue Langevin (ILL) France, using the Osaka crystal (mass 3.66 g). 
\par
For the LET experiment, the single crystal was aligned in the (0,K,L) horizontal scattering plane and a vertical field cryomagnet was used to apply a constant magnetic field of $B=4.7$~T along the ${\bf a}$-axis to reach the 1D QCP. These measurements were carried out at $T=0.3$~K using a $^3$He-insert. This temperature is well below the N\'eel temperature ($T_{\mathrm{N}}=5.5$~K) ensuring the presence of the effective longitudinal perturbing field necessary to stabilize $E_8$ physics. Using repetition rate multiplication and the chopper frequencies 280/140 Hz, incident neutron energies of $E_i$ = 22.69, 13.21, 8.51, 6.00, 4.42, 3.42, 2.70 meV were achieved with corresponding elastic energy resolutions of $\Delta E$ = 0.91, 0.41, 0.22, 0.14, 0.094, 0.065, 0.048 meV. The INS data were processed using the MANTID and HORACE software packages and converted to absolute units. 
\par
For the IN12 experiment, the crystal was aligned with the ${\bf a}$- and ${\bf c}$-axes within the horizontal instrumental scattering plane and a vertical DC magnetic field of 4.7 T was applied parallel to the ${\bf b}$-axis. A fixed final wavevector of $k_f = 1.15$~\AA$^{-1}$ was used, giving an energy resolution of $\Delta E \approx 0.114$~meV and wavevector resolution of $\approx 0.067$ r.l.u. A Beryllium filter was used to suppress higher-order wavelengths and spurious scattering. 

%$E_i$ = 22.69, 13.21, 8.51, 6.00, 4.42, 3.42, 2.70 meV were achieved with corresponding elastic energy resolutions of $\Delta E$ = 0.91, 0.41, 0.22, 0.14, 0.094, 0.065, 0.048 meV

\section{Theoretical model and dispersion of $E_8$ particles}

When applying a transverse field along (0,1,0) direction, the effective Hamiltonian for BaCo$_2$V$_2$O$_8$ is described by a 1D spin-1/2 Heisenberg-Ising model~\cite{Kimura_2013,Weiqiang2019,Zou_2021,Zou_2019}:
\begin{equation}
\begin{aligned}
 \mathcal{H} &=H_{XXZ}+H_{t}+H_{s}\\
H_{XXZ}&= J\sum_{n}[S^z_{n}S^z_{n+1}+\epsilon(S^x_{n}S^x_{n+1}+S^y_{n}S^y_{n+1})]\\
H_{t}&=-\mu_{B}g_{yy}H\sum_{n}[S^y_{n}+h_{x}(-1)^{n}S^{x}_{n}\\
& \;\;\;\; +h_{z}\cos(\pi\frac{2n-1}{4})S^z_{n}]\\
H_{s}&=-\mu_{B}H'\sum_{n}(-1)^n S^{z}_{n}
\label{eq:Hamil}
\end{aligned}
\end{equation}
where $S^{\alpha}_{n} = \frac{1}{2}\sigma^{\alpha}_{n}$($\alpha=x,y,z$) are spin-1/2 operators at site $n$ with Pauli matrices $\sigma^{\alpha}$. $J=5.8\;$meV, $\epsilon=0.46$, $h_{x(z)}=0.4(0.14)$, $g_{yy}=2.75$. The applied transverse field is set $\mu_{0}H=4.7\;$T, which is the critical field of the putative 1D QCP~\cite{Zou_2019,Zou_2021}. The effective staggered longitudinal field $\mu_{B}H'=0.018J$ comes from a mean-field treatment of the inter-chain coupling in the 3D ordering region below $T_{\mathrm{N}}$~\cite{Faure:2017iup,Zou_2021}.
 $H_s$ provides a necessary relevant perturbation for realizing the quantum $E_8$ physics \cite{Zou_2021}. Focusing on the parameter region around the putative 1D QCP, in the scaling limit, the effective Hamiltonian of the spin chain becomes~\cite{DELFINO1995724,Zou_2021,xiao_2021}
\begin{align}
\mathcal{H}_{E_8}=\mathcal{H}_{c=1/2}+h\int dx \sigma(x).
\label{eq:E8Hamil}
\end{align}
$\mathcal{H}_{c=1/2}$ is the Hamiltonian for a central charge 1/2 conformal
field theory, which describes the quantum critical physics of the TFIC.
$h$ and $\sigma (x)$ corresponding to the scaling limits of $\mu_{B}H'$ and $\sigma_j^z$
are the strengths of the perturbation field and the relevant primary field, respectively. To determine the dispersion of $E_8$ particles and compare with the spectrum measured by INS, we calculate the DSF in the field theory frame, $D^{\alpha\alpha}(\omega,q)=\sum_{n=1}^{\infty}\frac{(2\pi)^2}{\prod_{a_{i}=1}^{8} n_{a_{i}}!}\int_{-\infty}^{\infty}\prod_{j=1}^{n}\frac{d\theta_{j}}{2\pi}|\langle 0|\sigma^{\alpha}|A_{a_{1}}(\theta_{1})...A_{a_{n}}(\theta_{n})\rangle|^{2}$\\
$\delta(\omega-\sum_{j=1}^{n}E_{j})\delta(q-\sum_{j=1}^{n}P_{j}),$
where $\alpha=x,z$, and $a_{i}=1...8$ are quasi-particles obtained from the quantum $E_{8}$ integrable theory~\cite{Zamolodchikov:1989fp,DELFINO1995724,Zou_2021,xiao_2021}. $n_{a_{i}}$ is the number of particle $a_{i}$ involved in the corresponding channel. $E_{j}=m_{a_{j}}\cosh\theta_{j}$ and $P_{j}=m_{a_{j}}\sinh\theta_{j}$ are the energy and momentum of particle $a_{j}$ in terms of the rapidity $\theta$, respectively. The two Dirac $\delta$-functions reflect the energy and momentum conservation of the scattering. The DSF of $\sigma^{x,z}$ can be directly calculated from quantum $E_{8}$ integrable field theory~\cite{xiao_2021}, and the DSF of $\sigma^{y}$ can be obtained from DSF of $\sigma^{z}$~\cite{PhysRevLett.113.247201}. 
%The analytical result for the dispersion of the lightest three $E_8$ particles is shown in Fig.~\ref{fig:plot1}(a).
For a better comparison of the theoretical prediction from quantum $E_8$ field theory with the INS experimental result, two subtle issues are worth noting.

%\begin{figure}[h]
%\includegraphics[width=1.0\linewidth]{map_6meV.png}
%\caption{Magnetic Intensity of BaCo$_2$V$_2$O$_8$ measured in a transverse magnetic field of $\mu_{0}H=\mu_0 H_{\bot}^{c,1D}= 4.7$~T at $T=0.3$~K using the LET spectrometer. The data is displayed in absolute units as a function of wavevector ${\bf Q}$= (0,0,L) and energy, for neutron incident energy (a) $E_i$= 13.20 meV (integration range: $-1.3 \leq H \leq 1.3$ \& $-1.5 \leq K \leq 1.5$), (b) $E_i$= 8.51 meV (integration range: $-1.1 \leq H \leq 1.1$ \& $-1.6 \leq K \leq 1.6$) and (c) $E_i$= 6 meV (integration range: $-1.0 \leq H \leq 1.0$ \& $-2 \leq K \leq 2$).
%}
%\label{fig:plot_maps}
%\end{figure}

1. In the above field theory frame calculation, the speed of light is set as $c=1$. For the quantum $E_8$ model, as a massive relativistic quantum field theory, the dispersion of the $E_8$ particles follows the massive relativistic dispersion $E_i^2 = \Delta _i^2 + p_i^2{c^2}$, where $\Delta_i = m_i c^2$ and $p_i = m_i c$ with the ``rest mass" of the $i^{th}$ $E_8$ particle $m_i$ and the ``speed of light" $c$. When coming to a real material which actually is a lattice discrete in space, we need to re-scale the dispersion of the $E_8$ particles with the proper energy scale and length (momentum) scale serving as IR cutoffs. The theoretically expected energy peak corresponding to the lightest $E_{8}$ particle $m_{1}$ can be estimated from $E_{m_{1}}^{\text{theory}}=C_{\text{lattice}}H'^{8/15} \approx 1.2 \, \text{meV}$
\cite{Yang_2023}, where $C_{\text{lattice}} = 4.010\cdots$ is a modified constant for the lattice which originally comes from quantum $E_{8}$ field theory~\cite{E8_lattice}.
The value of $E_{m_{1}}^{\text{iTEBD}}$ matches the minimum gap
$\Delta_1 = 1.26 \,\text{meV}$ observed at the zone center (corresponding to zero transfer momentum), thus $\Delta_1$ can naturally serve as IR cutoff of the energy scale for the experimental data.
Since $\Delta_1 = m_1 c^2$ then we can pick up the corresponding IR momentum cutoff $p_1 = m_1 c$. By applying these two IR cutoffs scales we arrive at
\begin{equation}
\frac{{E_i^2}}{{{\Delta_{1}^{2}}}} = \frac{{\Delta _i^2}}{{\Delta_{1}^{2}}} + \frac{{p_i^2}}{(\Delta_{1}/c)^{2}}= \frac{{\Delta_i^2}}{{{{\Delta_1}^2}}} + \left( {\frac{{\hbar (L - 2)\pi /2d}}{{{m_1}c}}} \right)^2,
\label{eq:fit}
\end{equation}
where $p_i=\hbar(L - 2)\pi /2d$ is the momentum transfer with respect to ${\bf Q}$ = (0,0,2) and $d=8.4192/4=2.105$ is the nearest neighbor distance between Co$^{2+}$ ions projected onto the chain direction.
We need to determine the value of $c$ to obtain the IR cutoff for the momentum, whose value cannot be uniquely determined by the analytical theory but actually depends on the microscopic details of the material.

2. The four-fold periodicity of BaCo$_2$V$_2$O$_8$ leads to the sizable zone-folding effect of the experimental measurement, which makes the $E_{8}$ particles' dispersion shadowed by additional spectra. Such an effect cannot be obtained from the field theory calculation, instead, we need to go back to the original effective lattice model. By comparing spectra obtained from the lattice model and the field theory, the $E_{8}$ particles' dispersion will be extracted. To make these two subtle issues clear, we carry out iTEBD simulation for the effective Hamiltonian Eq.~(\ref{eq:Hamil})
with $J=1$, $\epsilon=0.47$, and critical field $\mu_B g_{yy}H=0.15$~\cite{Zou_2019,Zou_2021},

\begin{align}
\begin{split}
D_{\rm{lat}}^{\alpha\alpha}(\omega,q)=\frac{1}{N}&\sum_{j,j'=1}^{N}\exp\{-iq(j-j')\}\\
&\times\int_{-\infty}^{\infty}dt \exp(i\omega t)\langle S_{j}^{\alpha}(t)S_{j'}^{\alpha}(0)\rangle,
\label{eq:DSF}
\end{split}
\end{align}
with total number of lattice sites $N$ ($N \to \infty$ in iTEBD), and spin-1/2 operators $S^{\alpha},\alpha=x,y,z$. The procedure of the iTEBD calculation is as follows: 1. Generate a four-periodic ground state wave function of the effective Hamiltonian Eq.~(\ref{eq:Hamil}) with the parameters. The imaginary time-evolution is done
with fifth-order Trotter-Suzuki decomposition~\cite{Hatano2005}, where the imaginary time
slide is set as $d\tau=0.01$. The convergence condition is chosen as the
difference of the norm of singular values in the matrix product states being smaller than $10^{-12}$. The truncated dimension is chosen as $\chi=45$~\cite{PhysRevLett.98.070201,PhysRevA.79.043601}. 2. Calculate the DSF [Eq.~(\ref{eq:DSF})]
for $S^{x}$ and $S^{z}$, while the DSF of $S^{y}$ can be obtained from DSF of $S^{z}$ by using $D^{yy}(\omega,q)=\omega^2 D^{zz}(\omega,q)/(4J^2)$~\cite{PhysRevLett.113.247201}. For calculating this DSF with iTEBD algorithm, we first do real time and space propagation in Heisenberg picture, then by using Fourier transformation
we transform the obtained result into momentum and energy space
to obtain the final spectrum. The real time evolution is done
by a second order Trotter-Suzuki decomposition with $t=200,~dt=0.02$ for obtaining a relatively high accurate result near the Brillouin zone center. 3. A zone-folding effect is necessary to consider when obtaining the final spectrum for comparison with INS experimental results.

\bibliography{cite}

\begin{thebibliography}{27}%
\makeatletter
\providecommand \@ifxundefined [1]{%
 \@ifx{#1\undefined}
}%
\providecommand \@ifnum [1]{%
 \ifnum #1\expandafter \@firstoftwo
 \else \expandafter \@secondoftwo
 \fi
}%
\providecommand \@ifx [1]{%
 \ifx #1\expandafter \@firstoftwo
 \else \expandafter \@secondoftwo
 \fi
}%
\providecommand \natexlab [1]{#1}%
\providecommand \enquote  [1]{``#1''}%
\providecommand \bibnamefont  [1]{#1}%
\providecommand \bibfnamefont [1]{#1}%
\providecommand \citenamefont [1]{#1}%
\providecommand \href@noop [0]{\@secondoftwo}%
\providecommand \href [0]{\begingroup \@sanitize@url \@href}%
\providecommand \@href[1]{\@@startlink{#1}\@@href}%
\providecommand \@@href[1]{\endgroup#1\@@endlink}%
\providecommand \@sanitize@url [0]{\catcode `\\12\catcode `\$12\catcode
  `\&12\catcode `\#12\catcode `\^12\catcode `\_12\catcode `\%12\relax}%
\providecommand \@@startlink[1]{}%
\providecommand \@@endlink[0]{}%
\providecommand \url  [0]{\begingroup\@sanitize@url \@url }%
\providecommand \@url [1]{\endgroup\@href {#1}{\urlprefix }}%
\providecommand \urlprefix  [0]{URL }%
\providecommand \Eprint [0]{\href }%
\providecommand \doibase [0]{https://doi.org/}%
\providecommand \selectlanguage [0]{\@gobble}%
\providecommand \bibinfo  [0]{\@secondoftwo}%
\providecommand \bibfield  [0]{\@secondoftwo}%
\providecommand \translation [1]{[#1]}%
\providecommand \BibitemOpen [0]{}%
\providecommand \bibitemStop [0]{}%
\providecommand \bibitemNoStop [0]{.\EOS\space}%
\providecommand \EOS [0]{\spacefactor3000\relax}%
\providecommand \BibitemShut  [1]{\csname bibitem#1\endcsname}%
\let\auto@bib@innerbib\@empty
%</preamble>
\bibitem [{\citenamefont {Sachdev}(2011)}]{sachdev_2011}%
  \BibitemOpen
  \bibfield  {author} {\bibinfo {author} {\bibfnamefont {S.}~\bibnamefont
  {Sachdev}},\ }\href@noop {} {\emph {\bibinfo {title} {Quantum Phase
  Transitions}}}\ (\bibinfo  {publisher} {Cambridge University Press},\
  \bibinfo {address} {Cambridge, England},\ \bibinfo {year} {2011})\ pp.\
  \bibinfo {pages} {1--521}\BibitemShut {NoStop}%
\bibitem [{\citenamefont {Zamolodchikov}(1989)}]{Zamolodchikov:1989fp}%
  \BibitemOpen
  \bibfield  {author} {\bibinfo {author} {\bibfnamefont {A.~B.}\ \bibnamefont
  {Zamolodchikov}},\ }\href {https://doi.org/10.1142/S0217751X8900176X}
  {\bibfield  {journal} {\bibinfo  {journal} {Int. J. Mod. Phys. A}\ }\textbf
  {\bibinfo {volume} {4}},\ \bibinfo {pages} {4235} (\bibinfo {year}
  {1989})}\BibitemShut {NoStop}%
\bibitem [{\citenamefont {Dorey}(1997)}]{dorey1996}%
  \BibitemOpen
  \bibfield  {author} {\bibinfo {author} {\bibfnamefont {P.}~\bibnamefont
  {Dorey}},\ }\href@noop {} {\bibfield  {journal} {\bibinfo  {journal} {Lect.
  Notes Phys.}\ }\textbf {\bibinfo {volume} {498}},\ \bibinfo {pages} {85}
  (\bibinfo {year} {1997})}\BibitemShut {NoStop}%
\bibitem [{\citenamefont {Braden}\ \emph {et~al.}(1990)\citenamefont {Braden},
  \citenamefont {Corrigan}, \citenamefont {Dorey},\ and\ \citenamefont
  {Sasaki}}]{dorey1990}%
  \BibitemOpen
  \bibfield  {author} {\bibinfo {author} {\bibfnamefont {H.}~\bibnamefont
  {Braden}}, \bibinfo {author} {\bibfnamefont {E.}~\bibnamefont {Corrigan}},
  \bibinfo {author} {\bibfnamefont {P.}~\bibnamefont {Dorey}},\ and\ \bibinfo
  {author} {\bibfnamefont {R.}~\bibnamefont {Sasaki}},\ }\href@noop {}
  {\bibfield  {journal} {\bibinfo  {journal} {Nucl. Phys. B}\ }\textbf
  {\bibinfo {volume} {338}},\ \bibinfo {pages} {689 } (\bibinfo {year}
  {1990})}\BibitemShut {NoStop}%
\bibitem [{\citenamefont {Pfeuty}(1970)}]{Pfeuty}%
  \BibitemOpen
  \bibfield  {author} {\bibinfo {author} {\bibfnamefont {P.}~\bibnamefont
  {Pfeuty}},\ }\href@noop {} {\bibfield  {journal} {\bibinfo  {journal} {Ann.
  Phys.}\ }\textbf {\bibinfo {volume} {59}},\ \bibinfo {pages} {79} (\bibinfo
  {year} {1970})}\BibitemShut {NoStop}%
\bibitem [{\citenamefont {Boyanovsky}(1989)}]{Daniel}%
  \BibitemOpen
  \bibfield  {author} {\bibinfo {author} {\bibfnamefont {D.}~\bibnamefont
  {Boyanovsky}},\ }\href@noop {} {\bibfield  {journal} {\bibinfo  {journal}
  {Phys. Rev. B}\ }\textbf {\bibinfo {volume} {39}},\ \bibinfo {pages} {6744}
  (\bibinfo {year} {1989})}\BibitemShut {NoStop}%
\bibitem [{\citenamefont {Delfino}\ and\ \citenamefont
  {Mussardo}(1995)}]{DELFINO1995724}%
  \BibitemOpen
  \bibfield  {author} {\bibinfo {author} {\bibfnamefont {G.}~\bibnamefont
  {Delfino}}\ and\ \bibinfo {author} {\bibfnamefont {G.}~\bibnamefont
  {Mussardo}},\ }\href@noop {} {\bibfield  {journal} {\bibinfo  {journal}
  {Nucl. Phys. B}\ }\textbf {\bibinfo {volume} {455}},\ \bibinfo {pages} {724 }
  (\bibinfo {year} {1995})}\BibitemShut {NoStop}%
\bibitem [{\citenamefont {Zou}\ \emph {et~al.}(2021)\citenamefont {Zou},
  \citenamefont {Cui}, \citenamefont {Wang}, \citenamefont {Zhang},
  \citenamefont {Yang}, \citenamefont {Xu}, \citenamefont {Okutani},
  \citenamefont {Hagiwara}, \citenamefont {Matsuda}, \citenamefont {Wang},\
  and\ \citenamefont {et~al.}}]{Zou_2021}%
  \BibitemOpen
  \bibfield  {author} {\bibinfo {author} {\bibfnamefont {H.}~\bibnamefont
  {Zou}}, \bibinfo {author} {\bibfnamefont {Y.}~\bibnamefont {Cui}}, \bibinfo
  {author} {\bibfnamefont {X.}~\bibnamefont {Wang}}, \bibinfo {author}
  {\bibfnamefont {Z.}~\bibnamefont {Zhang}}, \bibinfo {author} {\bibfnamefont
  {J.}~\bibnamefont {Yang}}, \bibinfo {author} {\bibfnamefont {G.}~\bibnamefont
  {Xu}}, \bibinfo {author} {\bibfnamefont {A.}~\bibnamefont {Okutani}},
  \bibinfo {author} {\bibfnamefont {M.}~\bibnamefont {Hagiwara}}, \bibinfo
  {author} {\bibfnamefont {M.}~\bibnamefont {Matsuda}}, \bibinfo {author}
  {\bibfnamefont {G.}~\bibnamefont {Wang}},\ and\ \bibinfo {author}
  {\bibnamefont {et~al.}},\ }\href
  {http://dx.doi.org/10.1103/PhysRevLett.127.077201} {\bibfield  {journal}
  {\bibinfo  {journal} {Phys. Rev. Lett.}\ }\textbf {\bibinfo {volume} {127}}
  (\bibinfo {year} {2021})}\BibitemShut {NoStop}%
\bibitem [{\citenamefont {Zhang}\ \emph {et~al.}(2020)\citenamefont {Zhang},
  \citenamefont {Amelin}, \citenamefont {Wang}, \citenamefont {Zou},
  \citenamefont {Yang}, \citenamefont {Nagel}, \citenamefont {R\~o\ om},
  \citenamefont {Dey}, \citenamefont {Nugroho}, \citenamefont {Lorenz},
  \citenamefont {Wu},\ and\ \citenamefont {Wang}}]{PhysRevB.101.220411}%
  \BibitemOpen
  \bibfield  {author} {\bibinfo {author} {\bibfnamefont {Z.}~\bibnamefont
  {Zhang}}, \bibinfo {author} {\bibfnamefont {K.}~\bibnamefont {Amelin}},
  \bibinfo {author} {\bibfnamefont {X.}~\bibnamefont {Wang}}, \bibinfo {author}
  {\bibfnamefont {H.}~\bibnamefont {Zou}}, \bibinfo {author} {\bibfnamefont
  {J.}~\bibnamefont {Yang}}, \bibinfo {author} {\bibfnamefont {U.}~\bibnamefont
  {Nagel}}, \bibinfo {author} {\bibfnamefont {T.}~\bibnamefont {R\~o\ om}},
  \bibinfo {author} {\bibfnamefont {T.}~\bibnamefont {Dey}}, \bibinfo {author}
  {\bibfnamefont {A.~A.}\ \bibnamefont {Nugroho}}, \bibinfo {author}
  {\bibfnamefont {T.}~\bibnamefont {Lorenz}}, \bibinfo {author} {\bibfnamefont
  {J.}~\bibnamefont {Wu}},\ and\ \bibinfo {author} {\bibfnamefont
  {Z.}~\bibnamefont {Wang}},\ }\href
  {https://doi.org/10.1103/PhysRevB.101.220411} {\bibfield  {journal} {\bibinfo
   {journal} {Phys. Rev. B}\ }\textbf {\bibinfo {volume} {101}},\ \bibinfo
  {pages} {220411} (\bibinfo {year} {2020})}\BibitemShut {NoStop}%
\bibitem [{\citenamefont {Wang}\ \emph {et~al.}(2021)\citenamefont {Wang},
  \citenamefont {Zou}, \citenamefont {H\'ods\'agi}, \citenamefont {Kormos},
  \citenamefont {Tak\'acs},\ and\ \citenamefont {Wu}}]{xiao_2021}%
  \BibitemOpen
  \bibfield  {author} {\bibinfo {author} {\bibfnamefont {X.}~\bibnamefont
  {Wang}}, \bibinfo {author} {\bibfnamefont {H.}~\bibnamefont {Zou}}, \bibinfo
  {author} {\bibfnamefont {K.}~\bibnamefont {H\'ods\'agi}}, \bibinfo {author}
  {\bibfnamefont {M.}~\bibnamefont {Kormos}}, \bibinfo {author} {\bibfnamefont
  {G.}~\bibnamefont {Tak\'acs}},\ and\ \bibinfo {author} {\bibfnamefont
  {J.}~\bibnamefont {Wu}},\ }\href
  {https://doi.org/10.1103/PhysRevB.103.235117} {\bibfield  {journal} {\bibinfo
   {journal} {Phys. Rev. B}\ }\textbf {\bibinfo {volume} {103}},\ \bibinfo
  {pages} {235117} (\bibinfo {year} {2021})}\BibitemShut {NoStop}%
\bibitem [{\citenamefont {Coldea}\ \emph {et~al.}(2010)\citenamefont {Coldea},
  \citenamefont {Tennant}, \citenamefont {Wheeler}, \citenamefont {Wawrzynska},
  \citenamefont {Prabhakaran}, \citenamefont {Telling}, \citenamefont
  {Habicht}, \citenamefont {Smeibidl},\ and\ \citenamefont
  {Kiefer}}]{Coldea_2010}%
  \BibitemOpen
  \bibfield  {author} {\bibinfo {author} {\bibfnamefont {R.}~\bibnamefont
  {Coldea}}, \bibinfo {author} {\bibfnamefont {D.~A.}\ \bibnamefont {Tennant}},
  \bibinfo {author} {\bibfnamefont {E.~M.}\ \bibnamefont {Wheeler}}, \bibinfo
  {author} {\bibfnamefont {E.}~\bibnamefont {Wawrzynska}}, \bibinfo {author}
  {\bibfnamefont {D.}~\bibnamefont {Prabhakaran}}, \bibinfo {author}
  {\bibfnamefont {M.}~\bibnamefont {Telling}}, \bibinfo {author} {\bibfnamefont
  {K.}~\bibnamefont {Habicht}}, \bibinfo {author} {\bibfnamefont
  {P.}~\bibnamefont {Smeibidl}},\ and\ \bibinfo {author} {\bibfnamefont
  {K.}~\bibnamefont {Kiefer}},\ }\href
  {https://doi.org/10.1126/science.1180085} {\bibfield  {journal} {\bibinfo
  {journal} {Science}\ }\textbf {\bibinfo {volume} {327}},\ \bibinfo {pages}
  {177–180} (\bibinfo {year} {2010})}\BibitemShut {NoStop}%
\bibitem [{\citenamefont {Morris}\ \emph {et~al.}(2021)\citenamefont {Morris},
  \citenamefont {Desai}, \citenamefont {Viirok}, \citenamefont {Hüvonen},
  \citenamefont {Nagel}, \citenamefont {Rõõm}, \citenamefont {Krizan},
  \citenamefont {Cava}, \citenamefont {McQueen}, \citenamefont {Koohpayeh},
  \citenamefont {Kaul},\ and\ \citenamefont {Armitage}}]{Armitage_2021}%
  \BibitemOpen
  \bibfield  {author} {\bibinfo {author} {\bibfnamefont {C.~M.}\ \bibnamefont
  {Morris}}, \bibinfo {author} {\bibfnamefont {N.}~\bibnamefont {Desai}},
  \bibinfo {author} {\bibfnamefont {J.}~\bibnamefont {Viirok}}, \bibinfo
  {author} {\bibfnamefont {D.}~\bibnamefont {Hüvonen}}, \bibinfo {author}
  {\bibfnamefont {U.}~\bibnamefont {Nagel}}, \bibinfo {author} {\bibfnamefont
  {T.}~\bibnamefont {Rõõm}}, \bibinfo {author} {\bibfnamefont {J.~W.}\
  \bibnamefont {Krizan}}, \bibinfo {author} {\bibfnamefont {R.~J.}\
  \bibnamefont {Cava}}, \bibinfo {author} {\bibfnamefont {T.~M.}\ \bibnamefont
  {McQueen}}, \bibinfo {author} {\bibfnamefont {S.~M.}\ \bibnamefont
  {Koohpayeh}}, \bibinfo {author} {\bibfnamefont {R.~K.}\ \bibnamefont
  {Kaul}},\ and\ \bibinfo {author} {\bibfnamefont {N.~P.}\ \bibnamefont
  {Armitage}},\ }\href {https://doi.org/10.1038/s41567-021-01208-0} {\bibfield
  {journal} {\bibinfo  {journal} {Nature Phys.}\ }\textbf {\bibinfo {volume}
  {17}},\ \bibinfo {pages} {832–836} (\bibinfo {year} {2021})}\BibitemShut
  {NoStop}%
\bibitem [{\citenamefont {Fava}\ \emph {et~al.}(2020)\citenamefont {Fava},
  \citenamefont {Coldea},\ and\ \citenamefont {Parameswaran}}]{Fava25219}%
  \BibitemOpen
  \bibfield  {author} {\bibinfo {author} {\bibfnamefont {M.}~\bibnamefont
  {Fava}}, \bibinfo {author} {\bibfnamefont {R.}~\bibnamefont {Coldea}},\ and\
  \bibinfo {author} {\bibfnamefont {S.~A.}\ \bibnamefont {Parameswaran}},\
  }\href {https://doi.org/10.1073/pnas.2007986117} {\bibfield  {journal}
  {\bibinfo  {journal} {Proc. Nat. Acad. Sci.}\ }\textbf {\bibinfo {volume}
  {117}},\ \bibinfo {pages} {25219} (\bibinfo {year} {2020})}\BibitemShut
  {NoStop}%
\bibitem [{\citenamefont {Faure}\ \emph {et~al.}(2018)\citenamefont {Faure}
  \emph {et~al.}}]{Faure:2017iup}%
  \BibitemOpen
  \bibfield  {author} {\bibinfo {author} {\bibfnamefont {Q.}~\bibnamefont
  {Faure}} \emph {et~al.},\ }\href {https://doi.org/10.1038/s41567-018-0126-8}
  {\bibfield  {journal} {\bibinfo  {journal} {Nature Phys.}\ }\textbf {\bibinfo
  {volume} {14}},\ \bibinfo {pages} {716} (\bibinfo {year} {2018})}\BibitemShut
  {NoStop}%
\bibitem [{\citenamefont {Niesen}\ \emph {et~al.}(2013)\citenamefont {Niesen},
  \citenamefont {Kolland}, \citenamefont {Seher}, \citenamefont {Breunig},
  \citenamefont {Valldor}, \citenamefont {Braden}, \citenamefont {Grenier},\
  and\ \citenamefont {Lorenz}}]{PhysRevB.87.224413}%
  \BibitemOpen
  \bibfield  {author} {\bibinfo {author} {\bibfnamefont {S.~K.}\ \bibnamefont
  {Niesen}}, \bibinfo {author} {\bibfnamefont {G.}~\bibnamefont {Kolland}},
  \bibinfo {author} {\bibfnamefont {M.}~\bibnamefont {Seher}}, \bibinfo
  {author} {\bibfnamefont {O.}~\bibnamefont {Breunig}}, \bibinfo {author}
  {\bibfnamefont {M.}~\bibnamefont {Valldor}}, \bibinfo {author} {\bibfnamefont
  {M.}~\bibnamefont {Braden}}, \bibinfo {author} {\bibfnamefont
  {B.}~\bibnamefont {Grenier}},\ and\ \bibinfo {author} {\bibfnamefont
  {T.}~\bibnamefont {Lorenz}},\ }\href
  {https://doi.org/10.1103/PhysRevB.87.224413} {\bibfield  {journal} {\bibinfo
  {journal} {Phys. Rev. B}\ }\textbf {\bibinfo {volume} {87}},\ \bibinfo
  {pages} {224413} (\bibinfo {year} {2013})}\BibitemShut {NoStop}%
\bibitem [{\citenamefont {Can\'evet}\ \emph {et~al.}(2013)\citenamefont
  {Can\'evet}, \citenamefont {Grenier}, \citenamefont
  {Klanj\ifmmode~\check{s}\else \v{s}\fi{}ek}, \citenamefont {Berthier},
  \citenamefont {Horvati\ifmmode~\acute{c}\else \'{c}\fi{}}, \citenamefont
  {Simonet},\ and\ \citenamefont {Lejay}}]{PhysRevB.87.054408}%
  \BibitemOpen
  \bibfield  {author} {\bibinfo {author} {\bibfnamefont {E.}~\bibnamefont
  {Can\'evet}}, \bibinfo {author} {\bibfnamefont {B.}~\bibnamefont {Grenier}},
  \bibinfo {author} {\bibfnamefont {M.}~\bibnamefont
  {Klanj\ifmmode~\check{s}\else \v{s}\fi{}ek}}, \bibinfo {author}
  {\bibfnamefont {C.}~\bibnamefont {Berthier}}, \bibinfo {author}
  {\bibfnamefont {M.}~\bibnamefont {Horvati\ifmmode~\acute{c}\else
  \'{c}\fi{}}}, \bibinfo {author} {\bibfnamefont {V.}~\bibnamefont {Simonet}},\
  and\ \bibinfo {author} {\bibfnamefont {P.}~\bibnamefont {Lejay}},\ }\href
  {https://doi.org/10.1103/PhysRevB.87.054408} {\bibfield  {journal} {\bibinfo
  {journal} {Phys. Rev. B}\ }\textbf {\bibinfo {volume} {87}},\ \bibinfo
  {pages} {054408} (\bibinfo {year} {2013})}\BibitemShut {NoStop}%
\bibitem [{\citenamefont {Cui}\ \emph {et~al.}(2019)\citenamefont {Cui},
  \citenamefont {Zou}, \citenamefont {Xi}, \citenamefont {He}, \citenamefont
  {Yang}, \citenamefont {Shu}, \citenamefont {Zhang}, \citenamefont {Hu},
  \citenamefont {Chen}, \citenamefont {Yu}, \citenamefont {Wu},\ and\
  \citenamefont {Yu}}]{Weiqiang2019}%
  \BibitemOpen
  \bibfield  {author} {\bibinfo {author} {\bibfnamefont {Y.}~\bibnamefont
  {Cui}}, \bibinfo {author} {\bibfnamefont {H.}~\bibnamefont {Zou}}, \bibinfo
  {author} {\bibfnamefont {N.}~\bibnamefont {Xi}}, \bibinfo {author}
  {\bibfnamefont {Z.}~\bibnamefont {He}}, \bibinfo {author} {\bibfnamefont
  {Y.~X.}\ \bibnamefont {Yang}}, \bibinfo {author} {\bibfnamefont
  {L.}~\bibnamefont {Shu}}, \bibinfo {author} {\bibfnamefont {G.~H.}\
  \bibnamefont {Zhang}}, \bibinfo {author} {\bibfnamefont {Z.}~\bibnamefont
  {Hu}}, \bibinfo {author} {\bibfnamefont {T.}~\bibnamefont {Chen}}, \bibinfo
  {author} {\bibfnamefont {R.}~\bibnamefont {Yu}}, \bibinfo {author}
  {\bibfnamefont {J.}~\bibnamefont {Wu}},\ and\ \bibinfo {author}
  {\bibfnamefont {W.}~\bibnamefont {Yu}},\ }\href
  {https://doi.org/10.1103/PhysRevLett.123.067203} {\bibfield  {journal}
  {\bibinfo  {journal} {Phys. Rev. Lett.}\ }\textbf {\bibinfo {volume} {123}},\
  \bibinfo {pages} {067203} (\bibinfo {year} {2019})}\BibitemShut {NoStop}%
\bibitem [{\citenamefont {Vidal}(2007)}]{PhysRevLett.98.070201}%
  \BibitemOpen
  \bibfield  {author} {\bibinfo {author} {\bibfnamefont {G.}~\bibnamefont
  {Vidal}},\ }\href {https://doi.org/10.1103/PhysRevLett.98.070201} {\bibfield
  {journal} {\bibinfo  {journal} {Phys. Rev. Lett.}\ }\textbf {\bibinfo
  {volume} {98}},\ \bibinfo {pages} {070201} (\bibinfo {year}
  {2007})}\BibitemShut {NoStop}%
\bibitem [{\citenamefont {Lake}\ \emph {et~al.}(2022)\citenamefont {Lake},
  \citenamefont {Puzniak}, \citenamefont {Ma},\ and\ \citenamefont
  {Balz}}]{LET}%
  \BibitemOpen
  \bibfield  {author} {\bibinfo {author} {\bibfnamefont {B.}~\bibnamefont
  {Lake}}, \bibinfo {author} {\bibfnamefont {K.}~\bibnamefont {Puzniak}},
  \bibinfo {author} {\bibfnamefont {J.}~\bibnamefont {Ma}},\ and\ \bibinfo
  {author} {\bibfnamefont {C.}~\bibnamefont {Balz}},\ }\emph {\bibinfo {title}
  {Dispersion of $E_8$ particles in the spin-1/2 antiferromagnetic XXZ chain
  BaCo$_{2}$V$_{2}$O$_{8}$ in a transverse magnetic field}},\ \href
  {https://doi.org/https://doi.org/10.5286/ISIS.E.RB2210086} {Ph.D. thesis}
  (\bibinfo {year} {2022})\BibitemShut {NoStop}%
\bibitem [{\citenamefont {Kimura}\ \emph {et~al.}(2013)\citenamefont {Kimura},
  \citenamefont {Okunishi}, \citenamefont {Hagiwara}, \citenamefont {Kindo},
  \citenamefont {He}, \citenamefont {Taniyama}, \citenamefont {Itoh},
  \citenamefont {Koyama},\ and\ \citenamefont {Watanabe}}]{Kimura_2013}%
  \BibitemOpen
  \bibfield  {author} {\bibinfo {author} {\bibfnamefont {S.}~\bibnamefont
  {Kimura}}, \bibinfo {author} {\bibfnamefont {K.}~\bibnamefont {Okunishi}},
  \bibinfo {author} {\bibfnamefont {M.}~\bibnamefont {Hagiwara}}, \bibinfo
  {author} {\bibfnamefont {K.}~\bibnamefont {Kindo}}, \bibinfo {author}
  {\bibfnamefont {Z.}~\bibnamefont {He}}, \bibinfo {author} {\bibfnamefont
  {T.}~\bibnamefont {Taniyama}}, \bibinfo {author} {\bibfnamefont
  {M.}~\bibnamefont {Itoh}}, \bibinfo {author} {\bibfnamefont {K.}~\bibnamefont
  {Koyama}},\ and\ \bibinfo {author} {\bibfnamefont {K.}~\bibnamefont
  {Watanabe}},\ }\href {https://doi.org/10.7566/JPSJ.82.033706} {\bibfield
  {journal} {\bibinfo  {journal} {J. Phys. Soc. Jpn.}\ }\textbf {\bibinfo
  {volume} {82}},\ \bibinfo {pages} {033706} (\bibinfo {year}
  {2013})}\BibitemShut {NoStop}%
\bibitem [{\citenamefont {Zou}\ \emph {et~al.}(2019)\citenamefont {Zou},
  \citenamefont {Yu},\ and\ \citenamefont {Wu}}]{Zou_2019}%
  \BibitemOpen
  \bibfield  {author} {\bibinfo {author} {\bibfnamefont {H.}~\bibnamefont
  {Zou}}, \bibinfo {author} {\bibfnamefont {R.}~\bibnamefont {Yu}},\ and\
  \bibinfo {author} {\bibfnamefont {J.}~\bibnamefont {Wu}},\ }\href
  {https://doi.org/10.1088/1361-648x/ab4c71} {\bibfield  {journal} {\bibinfo
  {journal} {J. Phys.: Condens. Matt.}\ }\textbf {\bibinfo {volume} {32}},\
  \bibinfo {pages} {045602} (\bibinfo {year} {2019})}\BibitemShut {NoStop}%
\bibitem [{\citenamefont {Wu}\ \emph {et~al.}(2014)\citenamefont {Wu},
  \citenamefont {Kormos},\ and\ \citenamefont {Si}}]{PhysRevLett.113.247201}%
  \BibitemOpen
  \bibfield  {author} {\bibinfo {author} {\bibfnamefont {J.}~\bibnamefont
  {Wu}}, \bibinfo {author} {\bibfnamefont {M.}~\bibnamefont {Kormos}},\ and\
  \bibinfo {author} {\bibfnamefont {Q.}~\bibnamefont {Si}},\ }\href
  {http://dx.doi.org/10.1103/PhysRevLett.113.247201} {\bibfield  {journal}
  {\bibinfo  {journal} {Phys. Rev. Lett.}\ }\textbf {\bibinfo {volume} {113}},\
  \bibinfo {pages} {247201} (\bibinfo {year} {2014})}\BibitemShut {NoStop}%
\bibitem [{\citenamefont {Yang}\ \emph {et~al.}(2023)\citenamefont {Yang},
  \citenamefont {Wang},\ and\ \citenamefont {Wu}}]{Yang_2023}%
  \BibitemOpen
  \bibfield  {author} {\bibinfo {author} {\bibfnamefont {J.}~\bibnamefont
  {Yang}}, \bibinfo {author} {\bibfnamefont {X.}~\bibnamefont {Wang}},\ and\
  \bibinfo {author} {\bibfnamefont {J.}~\bibnamefont {Wu}},\ }\href
  {https://doi.org/10.1088/1751-8121/acad48} {\bibfield  {journal} {\bibinfo
  {journal} {J. Phys. A: Math. Theor.}\ }\textbf {\bibinfo {volume} {56}},\
  \bibinfo {pages} {013001} (\bibinfo {year} {2023})}\BibitemShut {NoStop}%
\bibitem [{\citenamefont {Caselle}\ and\ \citenamefont
  {Hasenbusch}(2000)}]{E8_lattice}%
  \BibitemOpen
  \bibfield  {author} {\bibinfo {author} {\bibfnamefont {M.}~\bibnamefont
  {Caselle}}\ and\ \bibinfo {author} {\bibfnamefont {M.}~\bibnamefont
  {Hasenbusch}},\ }\href
  {https://doi.org/https://doi.org/10.1016/S0550-3213(00)00074-2} {\bibfield
  {journal} {\bibinfo  {journal} {Nucl. Phys. B}\ }\textbf {\bibinfo {volume}
  {579}},\ \bibinfo {pages} {667} (\bibinfo {year} {2000})}\BibitemShut
  {NoStop}%
\bibitem [{\citenamefont {Hatano}\ and\ \citenamefont
  {Suzuki}(2005)}]{Hatano2005}%
  \BibitemOpen
  \bibfield  {author} {\bibinfo {author} {\bibfnamefont {N.}~\bibnamefont
  {Hatano}}\ and\ \bibinfo {author} {\bibfnamefont {M.}~\bibnamefont
  {Suzuki}},\ }\href {https://doi.org/10.1007/11526216_2} {\emph {\bibinfo
  {title} {Quantum Annealing and Other Optimization Methods}}}\ (\bibinfo
  {publisher} {Springer Berlin Heidelberg},\ \bibinfo {address} {Berlin,
  Heidelberg},\ \bibinfo {year} {2005})\ pp.\ \bibinfo {pages}
  {37--68}\BibitemShut {NoStop}%
\bibitem [{\citenamefont {Danshita}\ and\ \citenamefont
  {Naidon}(2009)}]{PhysRevA.79.043601}%
  \BibitemOpen
  \bibfield  {author} {\bibinfo {author} {\bibfnamefont {I.}~\bibnamefont
  {Danshita}}\ and\ \bibinfo {author} {\bibfnamefont {P.}~\bibnamefont
  {Naidon}},\ }\href {https://doi.org/10.1103/PhysRevA.79.043601} {\bibfield
  {journal} {\bibinfo  {journal} {Phys. Rev. A}\ }\textbf {\bibinfo {volume}
  {79}},\ \bibinfo {pages} {043601} (\bibinfo {year} {2009})}\BibitemShut
  {NoStop}%
\bibitem [{\citenamefont {Borthwick}\ and\ \citenamefont
  {Garibaldi}(2011)}]{borthwick2011did}%
  \BibitemOpen
  \bibfield  {author} {\bibinfo {author} {\bibfnamefont {D.}~\bibnamefont
  {Borthwick}}\ and\ \bibinfo {author} {\bibfnamefont {S.}~\bibnamefont
  {Garibaldi}},\ }\href@noop {} {\bibfield  {journal} {\bibinfo  {journal}
  {Not. Amer. Math. Soc.}\ }\textbf {\bibinfo {volume} {58}},\ \bibinfo {pages}
  {1055} (\bibinfo {year} {2011})}\BibitemShut {NoStop}%
\end{thebibliography}%
\bibliographystyle{apsrev4-2}

\end{document}